# Multilingual Ontology Matching based on Wiktionary Data Accessible via SPARQL Endpoint ♣

© Feiyu Lin

Jönköping University, Sweden
feiyu.lin@jth.hj.se

© Andrew Krizhanovsky

Institution of the Russian Academy of Sciences St.Petersburg Institute for Informatics and Automation RAS
andrew dot krizhanovsky@gmail.com

## Abstract

Interoperability is a feature required by the Semantic Web. It is provided by the ontology matching methods and algorithms. But now ontologies are presented not only in English, but in other languages as well. It is important to use an automatic translation for obtaining correct matching pairs in multilingual ontology matching. The translation into many languages could be based on the Google Translate API, the Wiktionary database, etc. From the point of view of the balance of presence of many languages, of manually crafted translations, of a huge size of a dictionary, the most promising resource is the Wiktionary. It is a collaborative project working on the same principles as the Wikipedia. The parser of the Wiktionary was developed and the machine-readable dictionary was designed. The data of the machine-readable Wiktionary are stored in a relational database, but with the help of D2R server the database is presented as an RDF store. Thus, it is possible to get lexicographic information (definitions, translations, synonyms) from web service using SPARQL requests. In the case study, the problem entity is a task of multilingual ontology matching based on Wiktionary data accessible via SPARQL endpoint. Ontology matching results obtained using Wiktionary were compared with results based on Google Translate API.

## 1 Introduction

Ontology matching is the process of finding correspondences between ontologies to allow them to interoperate. There are different methods, algorithms and systems designed for ontology matching [9].

A relatively new direction is concerned with an alignment of ontologies presented in different languages, i.e. multilingual ontology matching [8], [24]. There are different strategies related to multilingual ontology matching [24]: (1) the *indirect alignment* strategy based on composition of alignments, (2) the *direct matching* between two ontologies, i.e., without intermediary ontologies and with the help of external resources (translations). The latter strategy is used in this work.

The Ontology Alignment Evaluation Initiative (OAEI) [1] was launched in 2004 with the goal of estimating and comparing different techniques and systems related to ontology alignment. OAEI provides some multilingual datasets (ontologies and reference alignments), which were used in this work in order to evaluate the ontology matching system.

The multilingual ontology matching platform is presented in this work. COMS (Context-base Ontology Matching System) [15] implements the multilingual ontology matching based on Google Translate API and the data of the English Wiktionary and SPARQL technology.

The Wiktionary (www.wiktionary.org) is a multilingual and multifunctional dictionary. The Wiktionary contains not only word's definitions, semantically related words (synonyms, hypernyms, etc.), translations, but also the pronunciations (phonetic transcriptions, audio files), hyphenations, etymologies, quotations, parallel texts (quotations with translations), figures (which illustrate meaning of the words).

Wiktionary is popular since it is freely available and contains huge database of words with translations to many languages. The salient properties of the Wiktionary are the multilinguality, the size, and the speed of evolution. It is difficult to compare dictionaries with the Wiktionary, since data quickly become outdated. E.g. the PanDictionary was compared with the Wiktionary data obtained in the year 2008, when it has 403 413 translations [19]. Two years later, in 2010, the English Wiktionary contained twice as much translations (964 019). [2] So, the Wiktionary is permanently growing in number of entries and in the scope of languages. Now the English Wiktionary

---

[1] See http://oaei.ontologymatching.org
[2] See http://en.wiktionary.org/wiki/User:AKA_MBG/Statistics:Translations

contains entries in about 770 different languages. The Wiktionary data are used:
- In *machine translation* between Dutch and Afrikaans [21];
- In the *text parsing* system NULEX, where some Wiktionary data (verb tense) were integrated with WordNet and VerbNet [18];
- In a *speech recognition and speech synthesis* as a basis for the rapid pronunciation dictionary creation [10].

The Resource Description Framework is a data model for representing information about World Wide Web resources. SPARQL [1] is a query language for this data model. It is standardized by the World Wide Web Consortium. Now SPARQL is supported by most RDF triple store.

With the help of D2R server [4] the data extracted from the Wiktionary are presented in the form of RDF store. So, lexicographic information extracted from the Wiktionary is accessible by using SPARQL requests. In the case study, the problem entity is a task of multilingual ontology matching based on Wiktionary data accessible via SPARQL endpoint.

The next section describes system architecture consisting of the ontology matching system, Wiktionary relational database, D2R server and SPARQL client. Section 3 presents multilingual ontology matching experiments based on Wiktionary and Google Translate API. The discussion concludes the paper.

## 2 System architecture

In this section the developed platform will be described. The key components are a Wiktionary relational database, COMS ontology matching system [15], and D2R server which provides access to the machine-readable Wiktionary via SPARQL endpoint.

### 2.1 Machine-readable Wiktionary

There is an approach where the data are extracted from different types of wiki sites for the further processing and semantic search [21]. In that approach it was developed special services that export structured data into RDF/XML format. These services were designed and tailored to specific wiki engines (MediaWiki, DokuWiki).

Our work had the more modest goal of extracting data from only one type of wiki site (Wiktionary), moreover, only one Wiktionary language edition (English). The important fact is that Wiktionary entries have well-defined structure. However this structure is specified not at the level of MediaWiki, but at the level of texts of Wiktionary entries. Taking into account the structure of Wiktionary entry yield much more interesting information than just "structured data in RDF/XML format". The following data was extracted from the English and Russian Wiktionaries: definitions, thesaurus and translations. An example of data extracted from the "beautiful" English Wiktionary entry is presented in Fig. 1: the first meaning, semantic relations (synonyms and antonyms) and translations related to the first meaning.

The developed Wiktionary parser (wikt_parser) is one of several tools that parse Wiktionary data. Other tools include Zawilinski parser (Polish words in English Wiktionary) [14], JWKTL (the English and the German versions of Wiktionary)[3]. Our parser *wikt_parser* differs in two areas:
1. It requires that the XML dump to be initially loaded into the MySQL database;
2. It transforms the Wiktionary database into the machine-readable dictionary and saves it as a smaller database (MySQL or SQLite) for later use.

The parser source code and the database of the machine-readable Wiktionary are available at the project site.[4]

An automatic data extraction and a transformation of the Wiktionary data are explained in [13]. The Wiktionary database used in the experiments and an example of Wiktionary-based translations are described in the section "*3.1 Wiktionary Database and SPARQL queries*".

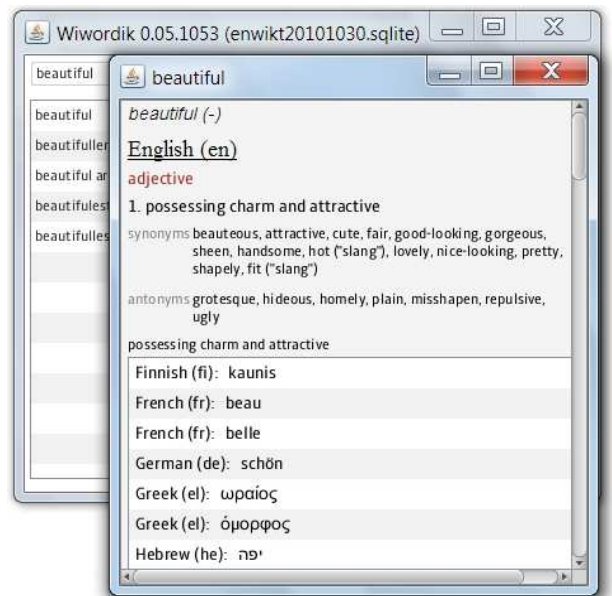

**Fig. 1.** An example of data extracted from the Wiktionary entry "beautiful": meaning (or definition), semantic relations (synonyms and antonyms) and translations of the first meaning.

### 2.2 COMS

COMS (Context-base Ontology Matching System) system consists of two parts: automatic ontology matching and context-based ontology matching [15]. The multilingual ontology matching is focus on automatic matching. Currently, COMS just finds the corresponding elements and presents the result as "elementA = elementB similarity measure (float)". The super, sub and inverse relationships are not included.

---
[3] See http://www.ukp.tu-darmstadt.de/software/jwktl/
[4] See http://code.google.com/p/wikokit/

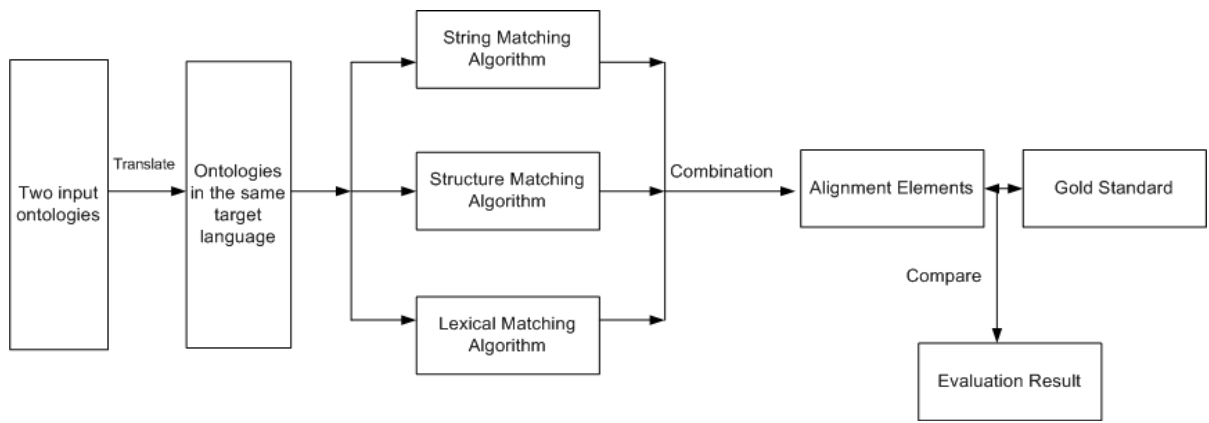

**Fig. 2.** An automatic ontology matching strategy and evaluation

The process of multilingual ontology matching involves two steps. First COMS translates the entities source to target ontology language. Then it applies automatically the following monolingual matching strategies. Fig. 2 shows two ontologies' automatic matching strategy and evaluation. Jena (http://jena.sourceforge.net) is used to parse ontology elements.

### 2.2.1 String Matching Strategy

Different string matching algorithms can be used here. There is a good survey [3] on the different string similarity methods to calculate string distance from *edit-distance* (e.g. Levenstein distance, Monger-Elkan distance, Jaro-Winkler distance) to *token-based distance* functions (e.g. Jaccard similarity, TF-IDF or cosine similarity, Jense-Shannon distance).

We use the Jaro-Winkler distance [25] and SmithWaterman algorithm [23] implemented in SimMetrics[5] and SecondString as our string matching methods. The threshold for Jaro-Winkler distance is 0.9. SmithWaterman algorithm can help find the similar region for two strings.

### 2.2.2 Structure Matching Strategy

Different structure matching strategies are implemented as following:
1. *If two elements of two ontologies' triples* (subject, predicate and object) are the same, the third element is assumed the same. For example, if the range and domain of two relations are the same, it means that the relations are the same. In future work, this will be extended to compare the common triples in the hierarchy.
2. *If the subclasses of two classes* are the same, these two classes are assumed the same. In future work, this will be extended to compare the common classes in the hierarchy.
3. *Expanding tree method* [17]. Ontology is expanded as a tree and set weights in the tree to calculate ontology concept similarity. The different levels are given different weights depending on the depth of the compared classes. The first level concepts, which get the weight as 3 are the class' subclasses and each relationship where it is domain or range. The second level concepts which get weight 2, are depending on the first level concepts' subclasses and their relationship's ranges. Similarity we can get the third level concepts, with weight 1, based on the second level concepts. We treat ontology matching as asymmetric. For example, a small ontology may perfectly match some parts of large ontology, the similarity between the small ontology and large ontology is 1.0 then, but not vice versa. The similarity between two concepts is computed as:

$$sim(x, y) = \frac{\sum w_{matched-concepts}}{\sum w_{x_i}}$$

### 2.2.3 Lexical Matching Strategy

One of our ontology matching strategies uses the WordNet (version 3.0). WordNet [6] is based on psycholinguistic theories to define word meaning and models not only word meaning associations but also meaning-meaning associations [7]. WordNet consists of a set of synsets. Synsets have different semantic relationships such as synonymy (similar) and antonymy (opposite), hypernymy (superconcept)/hyponymy (subconcept) (also called Is-A hierarchy / taxonomy), meronymy (part-of) and holonymy (has-a). The paper [16] provides an overview of how to apply WordNet in the ontology matching. In COMS, we use WordNet as the lexical dictionary.

---

[5] SimMetrics and SecondString are Java-based open-source packages used for string matching.

[6] http://wordnet.princeton.edu

WordNet-Similarity [7] has implemented several WordNet-based similarity measures in a Perl package. Java WordNet::Similarity[8] is a Java implementation of WordNet::Similarity. Jiang-Conrath [11] measure is chosen with threshold 1.0 to find corresponding classes in ontology matching. Jiang-Conrath measure is derived from the edge-based notion by adding the information content as a decision factor.

$$jcn = 1/(IC(synset1) + IC(synset2) - 2*IC(lcs)))$$

where *lcs* is the super concept of *synset1* and *synset2*, IC is the information content (of a synset).

For example, there are seven senses for the entry noun *school* hypernym relation in WordNet (fragment):

<u>Sense 1</u>

*school -- (an educational institution; "the school was founded in 1900")*

*=> educational institution -- (an institution dedicated to education)*

*=> institution, establishment -- (an organization founded and united for a specific purpose)*

*=> organization, organisation -- (a group of people who work together)*

*=> social group -- (people sharing some social relation)*

*=> group, grouping -- (any number of entities (members) considered as a unit)*

*=> abstraction, abstract entity -- (a general concept formed by extracting common features from specific examples)*

*=> entity -- (that which is perceived or known or inferred to have its own distinct existence (living or nonliving))*

<u>Sense 2</u>

*school, schoolhouse -- (a building where young people receive education; "the school was built in 1932"; "he walked to school every morning")*

*=> building, edifice -- (a structure that has a roof and walls and stands more or less permanently in one place; "there was a three-story building on the corner"; "it was an imposing edifice").*

Fig. 3 shows the fragment of nouns with *school* and *institution* in WordNet taxonomy. If *school* is used in Onto1 and *institution* is used in Onto2, *school* is the subconcept of *institution* in sense1 of WordNet. After we apply Jiang-Conrath measure, the similarity between *school* and *institution* is 1.25 that is bigger than threshold 1.0.

The following subsection describes how to access this database via SPARQL queries.

---

[7] http://www.d.umn.edu/~tpederse/similarity.html
[8] http://www.cogs.susx.ac.uk/users/drh21/

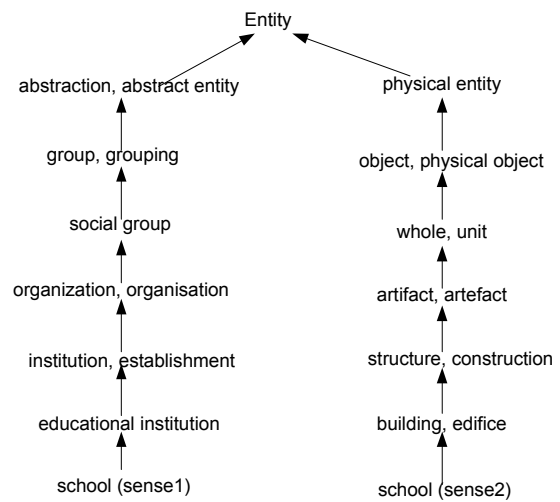

**Fig. 3.** The fragment of noun senses with *school* and *institution* in WordNet taxonomy

**2.3 SPARQL and D2RQ platform**

D2R server uses RDF and SPARQL languages in order to provide access to the relational database [4]. System takes SPARQL queries from the web and rewrites them to SQL queries via a specially prepared file (D2RQ mapping file).

The ontology matching system takes translation from the machine-readable Wiktionary with the help of D2R server (Fig. 4).

The D2RQ mapping file has to be created only once. After that it is possible to access to the relational database via SPARQL. SPARQL queries will be automatically translated on-the-fly into SQL by D2RQ platform. Therefore there is no need to replicate the database into RDF store.

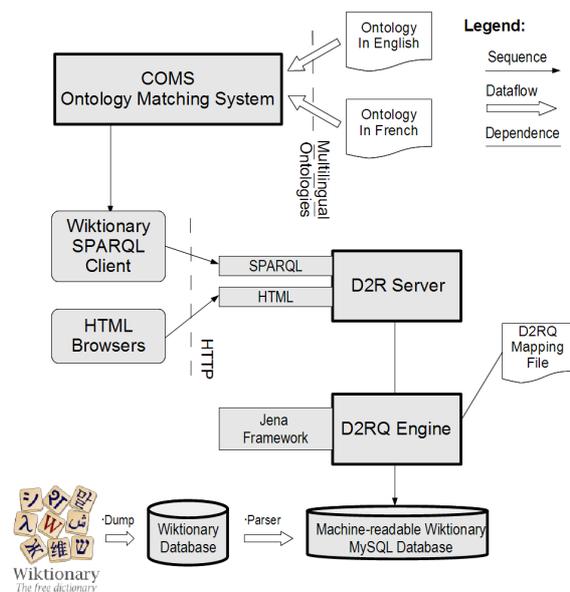

**Fig. 4.** Architecture of the platform integrating the ontology matching system with the machine-readable Wiktionary accessible via SPARQL queries

A simple Wiktionary SPARQL client was written in Java (as a part of COMS ontology matching system). It can obtain a list of translations from the source to the target language using Wiktionary data.

## 3 Experiments

The experiments are based on one benchmark track of OAEI. The reference ontology "test 101" is in English. This reference ontology contains 33 named classes, 24 object properties, 40 data properties, 56 named individuals and 20 anonymous individuals. The "test 206" of benchmark contains one ontology in French. Therefore one reference ontology (in English) is matched to French ontology.

In the "test 206" ontology in French the most part of words are presented in a canonical form (lemma). There are only a few words which are presented in non-canonical form, e.g. French words "articles", "auteurs", "éditeurs", "réalisateurs", "pages", "chapitres", "communications". Different word forms are recognized by the Google Translate system, but it is not taken into account by translation system based on the machine-readable Wiktionary.

Thus, our system translates labels from English to French first by using multilingual English Wiktionary, before applying monolingual matching procedures.

### 3.1 Wiktionary Database and SPARQL queries

The dump of the English Wiktionary (as of October 30, 2010) was the source data for our experiments. The created database of the machine-readable Wiktionary contains:
- 1 731 784 total entries;
- 269 405 English entries
- 154 990 French entries;
- 964 019 total number of translations;
- 50 617 number of translations from English to French.

This database was used for translation in the ontology matching system. This database was accessed via SPARQL queries. Most SPARQL queries are simple and short [2]. However, it turns out that it is not so in our case (Table 1).

Table 1 contains the example of the SPARQL request for the machine-readable Wiktionary. Input data for this request are (i) a language code (with value "en", i.e. English language), (ii) a Wiktionary entry ("rain cats and dogs"). Different colors of the rows in the Table 1 show different parts of the request, where one part corresponds to one table in the database.

The result of this request is translations of the English phrase "rain cats and dogs" into all languages presented in the Wiktionary. The part of the answer is presented in Table 2. Several SPARQL queries to the Wiktionary are presented on the wiki page of the project.[9]

---
[9] See http://code.google.com/p/wikokit/wiki/d2rqMappingSPARQL

**Table 1.** Sample SPARQL query for the machine-readable Wiktionary

```
SELECT ?langCode ?langName ?translationWord
WHERE {
  ?lang wikpa:lang_code "en";
        wikpa:lang_id ?langId.
  ?page wikpa:page_page_title "rain cats and dogs";
        wikpa:page_id ?pageId.
  ?lang_pos
        wikpa:lang_pos_page_id ?pageId;
        wikpa:lang_pos_lang_id ?langId;
        wikpa:lang_pos_id ?langPosId.
  ?meaning
        wikpa:meaning_id ?meaningId;
        wikpa:meaning_lang_pos_id ?langPosId.
  ?translation
        wikpa:translation_id ?translationId;
        wikpa:translation_lang_pos_id ?langPosId;
        wikpa:translation_meaning_id ?meaningId.
  ?langSource wikpa:lang_code ?langCode;
        wikpa:lang_name ?langName;
        wikpa:lang_id ?langIdSource.
  ?translation_entry
        wikpa:translation_entry_id ?translationEntryId;
        wikpa:translation_entry_translation_id ?translationId;
        wikpa:translation_entry_lang_id ?langIdSource;
        wikpa:translation_entry_wiki_text_id
  ?wikiTextIdTrans.
  ?wiki_text wikpa:wiki_text_id ?wikiTextIdTrans;
        wikpa:wiki_text_text ?translationWord.
} LIMIT 7
```

**Table 2.** SPARQL result: translations of the phrase "rain cats and dogs"

| ?langCode | ?langName | ?translationWord |
|---|---|---|
| cmn | Mandarin | 傾盆大雨 |
| cs | Czech | lít jako z konve |
| fr | French | pleuvoir des cordes |
| fr | French | pleuvoir à verse |
| fr | French | pleuvoir des hallebardes |
| ru | Russian | лить как из ведра |
| sv | Swedish | ösregna |

### 3.2 Translation Implementation

Ontology labels are often concatenated, e.g. "dateDePublication", "IntervalleDePages", "Extrait-Compilation". Google Translate system can recognize the label and translate directly. The machine-readable Wiktionary can't understand the concatenated label. In order to properly translate, labels are split into sequence of their constituent words. For example, "dateDePublication" is separated as "date De Publication".

In the reference alignment, one element is coming from "test 101" that is in English, one element is coming from "test 206" that is in French, and their similarity result. The total number of correct translations (the original French word and translated word compared to reference alignment) before applying ontology matching strategy by English Wiktionary is 44, and correct number by Google is 60.

The correct translation of English Wiktionary is lower, it is because that the Google gives the same word as translation if the word is not in the dictionary, e.g., "isbn", "url", "lccn", etc.. However, there is no translation in English Wiktionary to this case (see table 3, "isbn" example). On the other hand, Google translation API only provides one meaning translation of the words while English Wiktionary provides multiple meanings (if the word has) translation, for example, "Université" is only translated to "University" in Google, while is translated to "university; school" in English Wiktionary (see table 3). Google is good at translation the concatenated word, for example, "nomCourt" is translated "Shortname" directly (see table 3).

**Table 3.** Example of translations by Google and Wiktionary

| Source French word | Translation (list of words) | | Correspondence |
|---|---|---|---|
| Test #206 | By Google | By Wiktionary | Test 101 |
| Film | Film | movie; film; cinema; flick; motion picture | MotionPicture |
| Référence | Reference | | Reference |
| ExtraitLivre | BookExcerpt | | InBook |
| Partie | Party | part; subset; partially | Part |
| Livre | Paper | book; pound | Book |
| Conférence | Conference | lecture | Conference |
| Compilation | Compilation | | Collection |
| Université | University | university; school | School |
| isbn | isbn | | isbn |
| Clé | Key | key; radical; clef | key |
| nomCourt | Shortname | noun; name; short; court | shortName |
| dateDePub-lication | Publication Date | date; of; to; by; 's; in order to; publication; disclosure | firstPublished |
| chapitres | Chapters | | Chapters |
| éditeur | Editor | editor | Editor |

### 3.3 Precision and Recall

After we get the translation of the French ontology, COMS applies automatically the following monolingual matching strategies as described in Section 2.2. If there is no translation of the word, the original of element of the ontology is used to string matching, for example, "isbn" in Wiktionary case. Even COMS can get separate meaning of the concatenated word, but COMS doesn't support the different combination of the translation, for example, "nomCourt" is translated to "noun; name; short; court" and the correct translation is "shortName" (see table 3), COMS can't achieve to "shortName". "Film" is interpreted to "movie; film; cinema; flick; motion picture", COMS can recognize it is "MotionPicture".

The other matching strategies, such as WordNet is applied, e.g. "school" and "institution" similarity is 1.25 (see section 2.3). Structure matching strategy is applied, for example, in "Test 206", object property "articles" has domain "Revue" that interpreted as "Review" in Google and range "Article". In "Test 101" object property "articles" has domain "Journal" and range "Article". Since "articles" is similar "articles" and "Article" is similar "Article", even "Review" and "Journal" has no string similarity, based on structure similarity rules, "Revue" and "Journal" is similar. The final alignment result is based on the matching strategies presented in section 2.2.

There are different evaluation measures proposed in the OAEI, e.g., compliance and performance measures. The compliance measures consist of Precision, Recall, Fallout, F-measure, Overall, etc. Based on [6], the definition of precision and recall are:

**Definition (Precision).** *Given a reference alignment R, the precision of some alignment A is given by*

$$P(A,R) = \frac{R \cap A}{|A|}$$

It measures a valid possibility for ex post evaluations.

**Definition (Recall).** *Given a reference alignment R, the recall of some alignment A is given by*

$$R(A,R) = \frac{R \cap A}{|R|}$$

The provided reference alignment has 97 elements, which means $|R| = 97$.

The retrieved alignment based on English Wiktionary has 54 elements, which means $|A| = 54$, intersection $R \cap A = 53$

Precision is (see table 4):

$$P(A,R) = \frac{R \cap A}{|A|} = \frac{53}{54} = 0.98$$

Recall is (see table 4):

$$R(A,R) = \frac{R \cap A}{|R|} = \frac{53}{97} = 0.55$$

The retrieved alignment Google translation API of COMS has 61 elements, which means $|A|=61$, and intersection $R \cap A = 60$ elements.

Precision is (see table 4):

$$P(A,R) = \frac{R \cap A}{|A|} = \frac{60}{61} = 0.98$$

Recall is (see table 4):

$$R(A,R) = \frac{R \cap A}{|R|} = \frac{60}{97} = 0.62$$

**Table 4.** Precision and Recall Comparison between Wiktionary and Google

|  | Precision | Recall |
|---|---|---|
| **Wiktionary** | 0.98 | 0.55 |
| **Google** | 0.98 | 0.62 |

## 4 Discussion and conclusion

During the course of these investigations, the following problems were solved:
- the possibility to use the translation (extracted from the Wiktionary) via SPARQL endpoint was successfully verified;
- the different translation mechanisms (manually crafted Wiktionary translations and statistics-based Google translations) were applied and compared in ontology matching.

The using of SPARQL has cons and pros.

The merit of the approach (and SPARQL language in whole) is that the adding modifications to SPARQL request is simpler than the work with SQL request. This subjective point of view could be explained by the fact that one SPARQL request replaces many SQL requests (see Table 1). It allows more easily for going deep into details, since there is only one step between the question formulation and the result, i.e. there are no intermediate SQL requests.

The using of SPARQL has caveats and limitations though. A server which provides SPARQL endpoint could be easily broken or overloaded by poor, heavy or erroneous request [20]. The effective way is to constrain the infinity of SPARQL requests to a strictly defined set of functions in a web service. Thus SPARQL endpoint is necessary for developers and experimenters in order to check hypotheses, to create quickly complex queries. But the work should be carried out in a test mode, i.e. service can stop, fail, and it is possible to restart the service.

In order to improve results, the following problems should be solved:
1) Several Wiktionaries should be integrated into one machine-readable dictionary, since different Wiktionaries contains both overlapping and unique data (see the analysis of English Wiktionary and Russian Wiktionary in [12]).
2) Ontology context information should be taken into account (by integrating the translation process and the mapping activity [8], by using an information about a domain of the ontology [5]).
3) Now, in the experiment, (1) French words are translated into English, (2) monolingual matching procedures based on English WordNet were applied. There is an idea to use the free French WordNet [10] as an additional resource for the matching of two ontologies in English and French languages.

The Wiktionary parser development will be continued in future work, aiming at an extraction of quotes and Wiktionary context labels.

---

[10] See http://alpage.inria.fr/~sagot/wolf-en.html

♣ Part of this work was financed by the Foundation (The Swedish Institute), project CoReLib. Some parts of the research were carried out under projects funded by grants # 09-07-00066, # 09-07-00436 and # 11-01-00251 of the Russian Foundation for Basic Research, and project of the research program "Intelligent information technologies, mathematical modelling, system analysis and automation" of the Russian Academy of Sciences. This work is partly funded by the project DEON (Development and Evolution of Ontologies in Networked Organizations) based on a grant from STINT (The Swedish Foundation for International Cooperation in Research and Higher Education); grant IG 2008-2013.